# Mining Observation and Cognitive Behavior Process Patterns of Bridge Inspectors

Pengkun Liu, S.M.ASCE[1], Ruoxin Xiong, S.M.ASCE[2] and
Pingbo Tang, Ph.D., P.E., M.ASCE[3*]

[1] Dept. of Civil and Environmental Engineering, Carnegie Mellon University, 5000 Forbes Avenue, Pittsburgh, PA 15213; e-mail: pengkunl@andrew.cmu.edu
[2] Dept. of Civil and Environmental Engineering, Carnegie Mellon University, 5000 Forbes Avenue, Pittsburgh, PA 15213; e-mail: ruoxinx@andrew.cmu.edu
[3] Associate Professor, Dept. of Civil and Environmental Engineering, Carnegie Mellon University, 5000 Forbes Avenue, Pittsburgh, PA 15213; e-mail: ptang@andrew.cmu.edu

## ABSTRACT

In bridge inspection, engineers should diagnose the observed bridge defects by identifying factors underlying those defects. Traditionally, engineers search and organize structural condition-related information based on visual inspections and multiple data sources. Even following the same qualitative inspection standards, experienced engineers tend to find the critical defects and predict the underlying reasons more reliably than less experienced ones. Unique bridge and site conditions, quality of available data, and personal skills and knowledge collectively influence such a "subjective" nature of data-driven bridge diagnosis. Unfortunately, the lack of detailed data about how experienced engineers observe bridge defects and identify failure modes from multi-source data makes it hard to comprehend what engineers' behaviors form the best practice of producing reliable bridge inspection. Besides, even experienced engineers could sometimes fail in noticing critical defects, thereby producing inconsistent, conflicting condition assessments and biased maintenance plans. Therefore, detailed cognitive behavior analysis of bridge inspectors is critical for enabling a proactive inspector coaching system that uses many inspectors' behavior histories to complement personal limitations. This paper presents a computational framework for revealing engineers' observation and cognitive-behavioral processes to identify bridge defects and produce diagnosis conclusions based on observed defects. The authors designed a bridge inspection game consisting of FEM simulation data (stress and displacements) and inspection reports (basic bridge information and possible defect types) to capture and analyze experienced and inexperienced engineers' diagnosis behaviors. Mining these behavioral logs have revealed reusable behavioral process patterns that map critical bridge defects and diagnosis conclusions. The results indicate that the proposed method can proactively share inspection experiences and improve inspection processes' explainability and reliability.

## INTRODUCTION

Bridges are critical transportation infrastructures that serve daily life and socio-economic activities. It is significant to prioritize maintenance according to the bridges' damage severities with a limited maintenance budget. During bridge inspections, engineers should detect and explain the bridge defects to determine maintenance plans. Bridge engineers need to search and organize structural condition-related information based on visual inspections and multiple data sources for discovering bridge deterioration patterns during bridge inspection, which is tedious and subject to human reliability issues (Phares et al., 2004). Although many bridges have the same ratings, they may have different defects and reasons underlying the same condition rating numbers. Therefore, the processes of bridge inspection could hardly be explainable and have some limitations in guiding persuasive maintenance planning.

Engineer's observation and cognitive processes could explain how engineers capture bridge defects during the inspection and predict structural failure modes. A reliable diagnosis process will reveal all critical defects and correctly predict the bridges' failure modes. Because critical defects would be more specific than condition ratings, engineers' detailed cognitive diagnosis could produce more reliable and explainable outputs for maintenance planning. Such outputs should include a set of critical defects that capture structure damages and failure modes. Specifically, cognitive processes include mental processes, such as memory, learning, reasoning, and decision-making, to derive diagnosis outputs from a sequence of field observations (Smith and Kelly, 2015). Personal knowledge, experiences, and field conditions can collectively influence the observation and cognitive processes' reliability. When the field conditions are unfavorable and the amounts of field data are significant, engineers may sometimes lose focus among data, disregard some defects and produce unreliable conclusions about critical defects and failure modes. For example, a 40-year unnoticed defect about gusset plates bowing under stress is the Mississippi River Bridge Collapse's root cause (Williams, 2008). In 2003 even an engineer took a photo of this defect. The diagnosis process of groups of decision-makers underestimates this defect's importance. Therefore, engineers' underlying observation and cognitive behaviors using field data seem problematic and influence inspection reliability.

Existing works are limited but show some potential for using eye-tracking techniques to understand engineers' observation and cognitive behaviors. For example, the human eye movements, such as fixation count, total fixation time, and fixation duration captured through the eye-tracking technology, could infer humans' detailed observation and cognitive processes (Wang et al., 2018, Liu and Heynderickx, 2011). Unfortunately, limited studies examined engineers' detailed observation and cognitive processes deriving critical defects and underlying reasons from field conditions and available bridge project data.

Under the assumption that bridge engineers' experiences are associated with their cognitive behaviors, this paper focuses on studying how bridge engineers sequentially check various defects to identify critical defects and damages from different forms and contents of information delivery (e.g., reports and 3D models). The authors propose to collect and analyze experienced and inexperienced engineers' observation and cognitive processes. The purpose is to enable the discoveries of reusable observation and cognitive processes that can guide bridge inspectors in improving the explainability and reliability of inspection processes based on historical human behaviors. The following sections first present a motivating case, the research methodology, and this approach's experiment results and discussions.

**MOTIVATION CASE**

To understand the process of bridge inspections from observations to diagnosis, take a continuous rigid frame bridge (CRFB) as the case study (Figure 1). The routine inspection aims to evaluate, report the bridge's structural conditions, and infer corresponding reasons. Firstly, engineers inspect the bridge's appearance, focusing on the box girders, decks, and piers' defects. Then, engineers accurately select the bridge elements' critical defects, draw the defect distribution maps, and analyze the underlying causes.

As shown in Figure 1, the bridge has a total length of 1,010 m, including a six-span main bridge of 612m (66m + 4*120m + 66m) (Sun et al., 2020). Bridge engineers utilized the laser scanner and camera to get the bridge's point clouds and defect images to record the bridge's structural conditions. After that, with the finite element model (FEM) analysis, the engineers should infer the reasons behind the critical defects and write a conclusion section of the inspection report. Therefore, the diagnosis outputs should include critical defects and the corresponding reasons that capture the structure's underlying failure modes. For example, bridge engineers discover defects on the box girder in the observation stage. More specifically, there are cavities, transverse, and longitudinal cracks on the bottom slab and diagonal cracks of the web of the box girder. According to the FEM analysis of the bridge and the bridge engineers' knowledge, the reasons that cause such defects could be (1) tensile stress and (2) shearing stress.

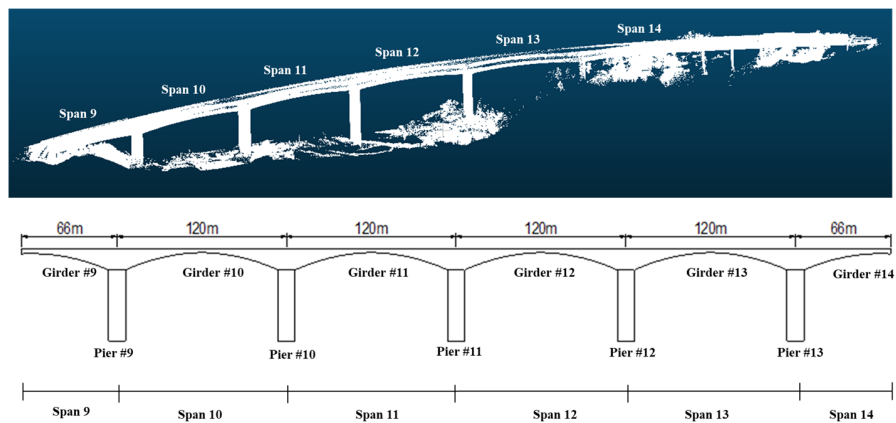

**Figure 1. Overview of the continuous rigid frame bridge.**

**METHODOLOGY**

This study aims to design a framework and harvest bridge engineers' cognitive behaviors to reveal engineers' experiences in bridge inspection. The authors developed a bridge inspection game that allows "players" (bridge inspectors) to assess the bridge defects and underlying reasons based on FEM simulation data (stress and displacements) and inspection reports (basic bridge information and possible defect types). During the game, the "players" will go through every bridge element and check various defects to identify critical defects within a time limit. The eye-tracking device and defect diagnosis table captured the "players" cognitive behaviors of identifying bridge defects and producing diagnosis conclusions based on observed defects, including eye movements, the sequences of the selections, timestamps, and the diagnosis results. Figure 2 shows the experimental flow consisting of bridge inspection environment preparation, bridge inspection environment introduction, the cognitive process of bridge inspection and diagnosis, and cognitive-behavioral data processing and analysis. This study recruited two graduate students from different majors (one from civil engineering with experience in bridge inspection and the other from transportation engineering with basic knowledge of bridge inspection).

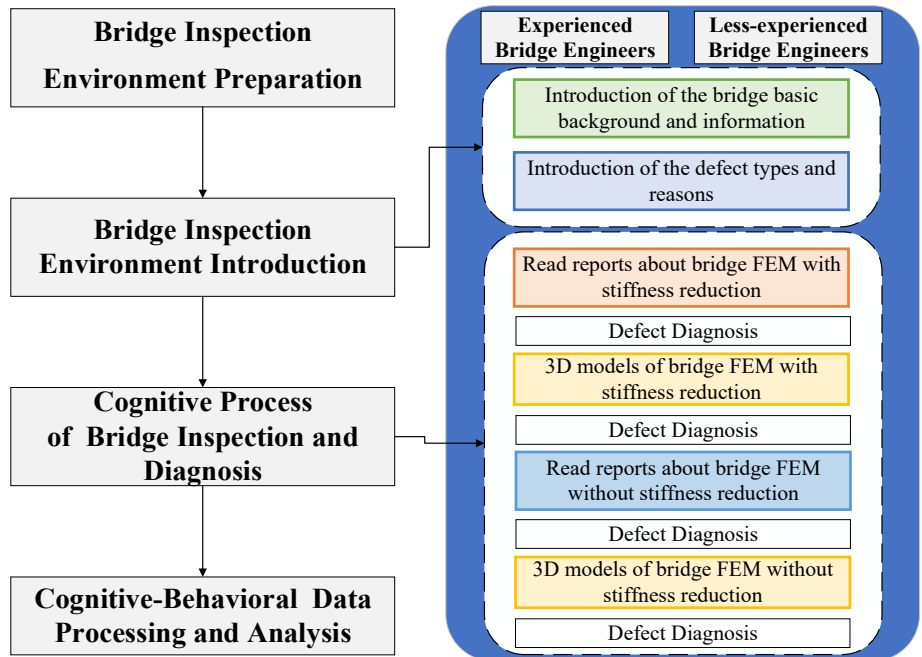

**Figure 2. Experimental flow.**

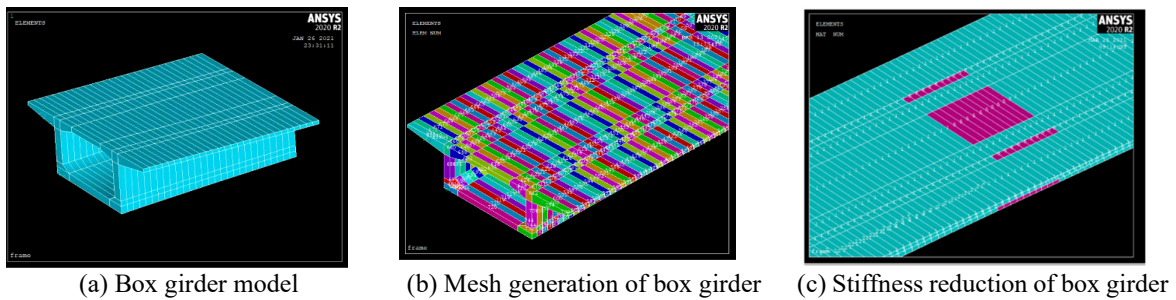

(a) Box girder model  (b) Mesh generation of box girder  (c) Stiffness reduction of box girder

**Figure 3. Defective bridge simulation results of a box girder by FEM analysis.**

**Bridge inspection environment preparation.** To secure the ground truths of defect types, locations, and the underlying reasons, the authors used FEM for simulating defective structures. We created FEM to simulate the defect developments on the box girders' top and bottom slabs in CRFBs through the stiffness reduction of those elements within FEM, as shown in Figure 3. Furthermore, the authors designed two FEMs of CRFBs to compare the effects of stiffness reduction. One model is the original CRFB model without any stiffness reduction (Figure 4). The other is the CRFB model with the stiffness reduction of the box-girders in mid-span 10, 11, 12 13 for simulating damages (Figure 5). One observation is that the stress distributions were significantly different between the two models, especially in Figure 4 (e), (f), and Figure 5 (e), (f).

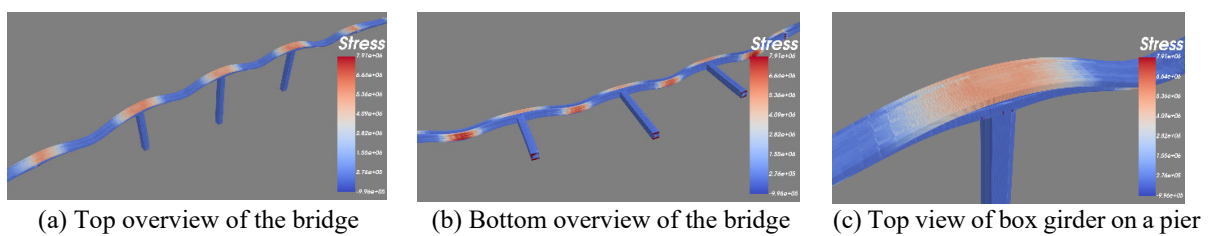

(a) Top overview of the bridge  (b) Bottom overview of the bridge  (c) Top view of box girder on a pier

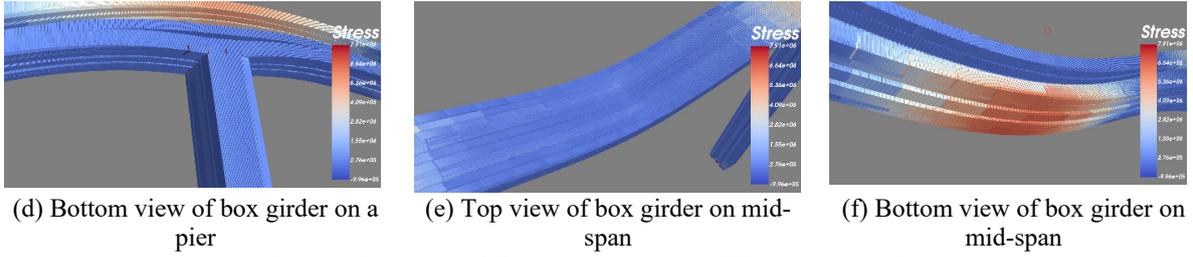

(d) Bottom view of box girder on a pier  (e) Top view of box girder on mid-span  (f) Bottom view of box girder on mid-span

**Figure 4. FEM of CRFB without stiffness reduction.**

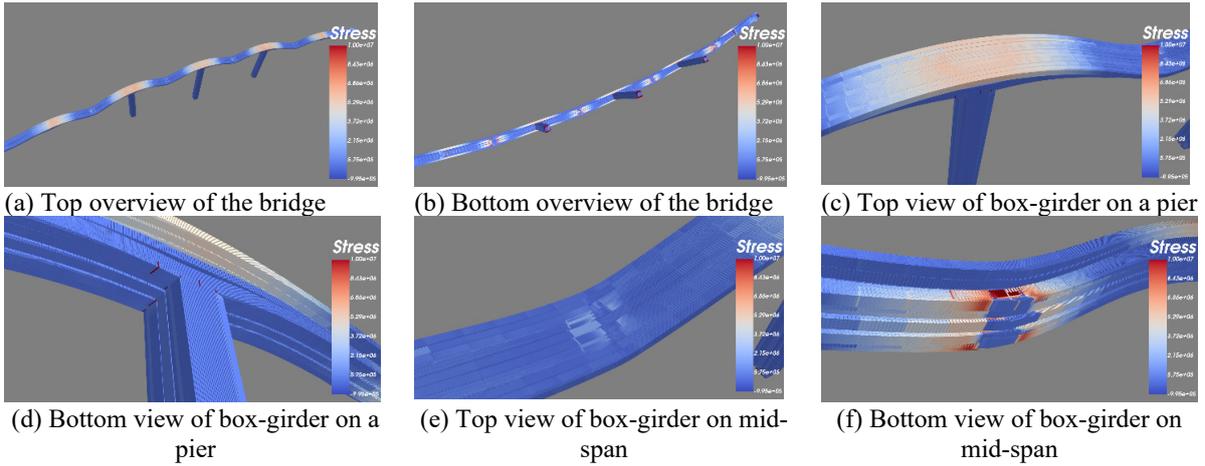

(a) Top overview of the bridge  (b) Bottom overview of the bridge  (c) Top view of box-girder on a pier

(d) Bottom view of box-girder on a pier  (e) Top view of box-girder on mid-span  (f) Bottom view of box-girder on mid-span

**Figure 5. FEM of CRFB with stiffness reductions on box-girders of mid-span 10, 11, 12, 13.**

**Bridge inspection environment introduction.** The participants can get familiar with the essential bridge background, possible defect types, and reasons before the experiment. The participants must find the critical defects, identify defect types and locations and write down the underlying reasons in the defect diagnosis stages after reading the reports with FEM images or interacting with the 3D models, as shown in Figure 6.

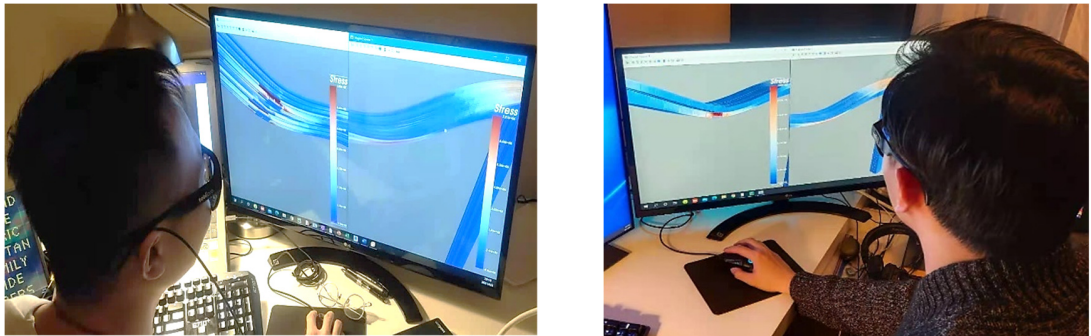

**Figure 6. Bridge inspection game with eye-tracking.**

**The cognitive process of diagnosis.** Once familiarized with the basic information and the required tasks, the participants commence the tasks. The participants can freely carry out their information searching in four types of materials about the CRFB, including the reports or models. Firstly, the participants read reports about bridge FEM with stiffness reduction and carry out a preliminary defect diagnosis. Then the participants interact with 3D CRFB FEM with stiffness reduction and implement a second defect diagnosis. Furthermore, participants read reports about bridge FEM without stiffness reduction and carry out a third defect diagnosis. Finally, participants interact with 3D CRFB FEM without stiffness reduction and implement a fourth defect diagnosis.

**EXPERIMENTAL RESULTS AND DISCUSSIONS**

**Table 1.** Experimental diagnosis results of participants

| Participant | First Diagnosis Reports with stiffness reduction | Second Diagnosis 3D models with stiffness reduction | Third Diagnosis Reports without stiffness reduction | Fourth Diagnosis 3D models without stiffness reduction |
|---|---|---|---|---|
| #1 Experienced | The bottom slab of the box girder in the middle of spans 10, 11, 12, 13 (Tension) Web of box girder in the middle of spans 10, 11, 12, 13 (Shear) | The bottom slab of the box girder in the middle of spans 10, 11, 12, 13 (Tension) Web of box girder in the middle of spans 10, 11, 12, 13 (Shear) The top slab of the box girder in the middle of spans 10, 11, 12, 13 (Tension) | The same | The same |
| #2 Inexperienced | Bottom of mid-span 10, 11,12 and 13 (Torsion) Piers of 9, 10, 11, 12, and 13 (Bending) Web of mid-spans 10, 11, 12, and 13 (Shear) | Bottom of mid-spans 10, 11,12 and 13 (Torsion) Piers of 9, 10, 11, 12, and 13 (Bending) Web of mid-spans 10, 11, 12, and 13 (Shear) Top of mid-spans 10, 11,12 and 13 (Tension) | Bottom of mid-spans 10, 11,12 and 13 (Torsion) Piers of 9, 10, 11, 12, and 13 (Bending) Web of mid-spans 10, 11, 12, and 13 (Shear) Top of mid-spans 10, 11, 12 and 13 (Tension) Bottom of mid-spans 10, 11,12 and 13 (Tension) | The same |

Mining the cognitive behaviors during bridge inspections could reveal reusable cognitive process patterns that map bridge defects and diagnosis conclusions. The authors used the diagnosis and related eye movements to assess the bridge engineers' cognitive progress, as shown in Table 1 and Figure 7. The authors find that significant differences between the two engineers are their experiences of the original CRFB's conditions. More precisely, experienced engineer #1 could accurately derive the actual conditions of CRFB without stiffness reductions. The experienced engineer could pay more attention to the critical locations where the stiffnesses were reduced and made a consistent diagnosis in the four diagnoses. Due to the stiffness reduction of the box girder's top and bottom slab, the shear stress on the box girder web increased, as shown in Figures 5 (e) and (f). The inexperienced engineer had no prior knowledge about the original CRFB's conditions. Therefore, the inexperienced engineer's policy was randomly searching for the defects, especially in the first and second diagnoses finding the apparent defects. However, once reading the reports without stiffness reduction, the inexperienced engineer knew the stress and deformation of the CRFB without defects. Then the inexperienced engineer could find parts of the critical defects, such as defects happening on the bottoms of mid-span 10, 11,12, and 13 because of the tension stress. Therefore, comparing the diagnosis results of the critical defects by interacting with different forms and contents of information delivery (reports or models with or without stiffness reduction), the authors find that the inexperienced engineers could find the critical defects by providing the CRFB baseline without stiffness reduction. Therefore, the CRFB baseline would be critical information that helps make a reliable and explainable inspection.

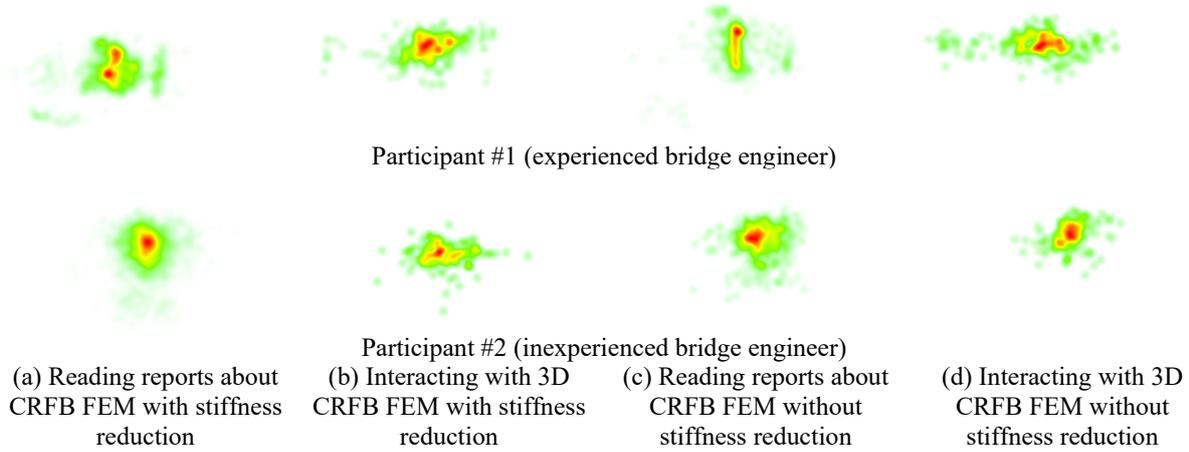

(a) Reading reports about CRFB FEM with stiffness reduction  (b) Interacting with 3D CRFB FEM with stiffness reduction  (c) Reading reports about CRFB FEM without stiffness reduction  (d) Interacting with 3D CRFB FEM without stiffness reduction

**Figure 7. Heat map of the participant gazes with four types of materials.**

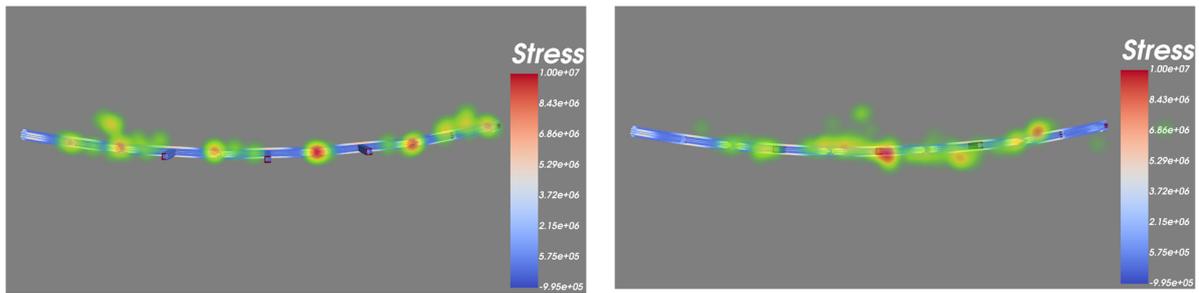

Participant #1 (experienced bridge engineer)   Participant #2 (inexperienced bridge engineer)

**Figure 8. Heat map of the participant gazes for finding the defects on bottom slabs.**

As shown in Figure 7, the engineers' gaze movements also infer their searching strategy and cognitive progress during the bridge inspection. As for the tasks of reading reports about CRFB FEM with or without stiffness reduction, as shown in Figure 7 (a) and (c), the gaze movements of the two participants all showed a vertical elongation pattern because the reports consisting of multi-pages, engineers need to scroll up and down to find defects. Besides, when interacting with 3D CRFB FEMs with or without stiffness reduction, as shown in Figures 7 (b) and (e), the engineers tended to search between the different spans. Therefore, the trajectory of the gaze movements shows a horizontal elongation pattern. The authors also identify apparent differences in the gaze movements between experienced and inexperienced engineers. Because of the random searching policy of the inexperienced engineer #2, the average trajectory of the gaze movements is more clustered and spherical compared to the experienced engineers. More precisely, as shown in Figure 8, in the tasks of finding the defects on the bottom of the bridge, the experienced bridge engineer could pay more attention to the bottom slab of each mid-span. In contrast, the inexperienced bridge engineer uses a random searching policy. Furthermore, according to the counts of eye movements, the inexperienced engineer would spend more time searching for defects, especially in the first task of reading reports about CRFB FEM with stiffness reduction.

## CONCLUSION AND FUTURE WORK

Capturing and analyzing cognitive processes about how engineers sequentially check various defects to find critical defects and predict the underlying reasons could reveal the patterns for keeping consistent inspection records. This study introduces a framework for revealing engineers' observation and cognitive processes through eye-tracking technology. The authors developed a bridge inspection game with reports and 3D models with or without stiffness reduction to track engineers' cognitive behaviors and defect diagnosis performances. Finally, the authors compared the cognitive progress of experienced and inexperienced engineers. The results indicate that the significant differences between the experienced and inexperienced engineers are prior knowledge about the original CRFB conditions without stiffness reduction. Once provided with the CRFB's stress and deformation baseline, the inexperienced engineer could find the critical defects. Therefore, the proposed method can proactively share inspection experiences and improve inspection processes' explainability and reliability. Finally, the authors identified a few limitations of the presented study. First, carry out more experiments to collect data from engineers of different backgrounds to prove the cognitive process differences. Second, include more detailed task-level inspection actions such as the click, rotation, and element selection sequence in future work. Third, mine the progress patterns from the detailed actions to build a participant-specific process model.

## ACKNOWLEDGEMENT

This material is based on work supported by the U.S. National Science Foundation (NSF) CAREER under Grant No. 1454654 and the U.S. National Science Foundation (NSF) Convergence under Grant No. 1937115. The authors gratefully acknowledged the supports.